# Three-dimensional Printing of Mycelium Hydrogels into Living Complex Materials

Silvan Gantenbein, Emanuele Colucci, Julian Käch, Etienne Trachsel, Fergal B. Coulter, Patrick A. Rühs, Kunal Masania[1]*, André R. Studart *

Complex Materials, Department of Materials, ETH Zürich, 8093 Zürich, Switzerland.

[1]Present address: Shaping Matter Lab, Faculty of Aerospace Engineering, Delft University of Technology, 2629 HS Delft, Netherlands.

*corresponding authors: K.Masania@tudelft.nl, andre.studart@mat.ethz.ch

**Abstract:**

Biological living materials, such as animal bones and plant stems, are able to self-heal, regenerate, adapt and make decisions under environmental pressures. Despite recent successful efforts to imbue synthetic materials with some of these remarkable functionalities, many emerging properties of complex adaptive systems found in biology remain unexplored in engineered living materials. Here, we report on a three-dimensional printing approach that harnesses the emerging properties of fungal mycelium to create living complex materials that self-repair, regenerate and adapt to the environment while fulfilling an engineering function. Hydrogels loaded with the fungus *Ganoderma lucidum* are 3D printed into lattice architectures to enable mycelial growth in a balanced exploration and exploitation pattern that simultaneously promotes colonization of the gel and bridging of air gaps. To illustrate the potential of such living complex materials, we 3D print a robotic skin that is mechanically robust, self-cleaning, and able to autonomously regenerate after damage.

**Main Text:**

Living organisms have long been exploited to synthesize materials for engineering needs, such as silk, cellulose and wood (*1, 2*), and have also led to materials with enhanced functionalities (*3*). This strategy has gained new impulse in recent years motivated by the sustainable nature of biological processes and the possibility of creating synthetic materials with living functionalities through the incorporation of microorganisms in abiotic host matrices (*4-10*). The combination of material-producing microorganisms with abiotic synthetic building blocks has led to the development of self-healing concretes for construction (*4, 11*), antibiotic-releasing living surfaces (*6*), genetically programmable living materials for sensing (*5, 7*) and biomedical applications (*8*), as well as functional living inks for 3D printing of bacteria-laden materials (*9, 12*). The ability to program microorganisms using synthetic biological tools (*2, 7, 8*) and to 3D print living inks into intricate geometries (*7, 9*) potentially designed by artificial intelligence (*13*) open enticing perspectives in the growing field of engineered living materials (*14, 15*).

Among the different microorganisms exploited so far, fungal mycelium is particularly interesting because it is a living complex adaptive system with emergent collective properties (*16-20*). Such complex behavior results from the fact that fungal mycelia have evolved to explore and exploit environments with distributed patches of nutrients either as decomposers or as symbionts with plants. To optimally gather these nutrients, mycelia form large networks consisting of interconnected elongated cells called hypha. The hyphae locally absorb water and nutrients and use it to power mycelial exploration of the surrounding or the formation of fruiting bodies for even further dissemination. These features make mycelium a fascinating biological example of a complex adaptive system comprising a network of decentralized units (cells) that self-organize into hierarchical structures with emerging collective behavior (*16*). In fungal mycelia, adaptation to the environment is manifested by its ability to change growth patterns between exploratory and exploitation strategies according to the local availability of nutrients. Such adaptive behavior gives rise to growth, communication and decision-making properties that are highly desired in synthetic materials.

Several approaches have been pursued to exploit mycelium-based materials for engineering purposes. Mycelium grown on agricultural waste acts as a binder to create sustainable materials that have been commercialized as environmental-friendly boards for insulation and packaging (*21*). Mycelial structures have also been fabricated for architectural applications (*22*), as leather-like self-grown films with tunable mechanical properties (*23, 24*) or as biological template to synthesize inorganic filaments (*25*). However, the fungi utilized for the fabrication of all these materials and products eventually die at the end of the process or are removed from the structure, which thereby do not benefit from the unique adaptive living properties of the microorganism.

Here, we combine the emerging adaptive behavior of mycelial networks with the capability of shaping matter in three dimensions to create living complex materials that serve specific engineering purposes. Because the living organism has evolved for a biological function, harnessing its emerging properties for other applications requires one to shape them in new structures. To address this, we print open architectures that provide a 3D environment for the growth of mycelia outside the host hydrogel into a functional complex material. The amount of nutrients available is deliberately controlled to enable the living material to find the right exploration-exploitation balance needed for the mycelium to grow, self-regenerate and autonomously adapt in response to the environment.

Living complex materials were 3D printed using hydrogels inoculated with mycelium. The workflow involves mixing of an agar-based hydrogel containing malt extract and rheology modifiers to obtain a granulated ink, the deposition of the fungus on top of such granular inks, and finally the removal of the fungus top layer to create the mycelium-laden feedstock for 3D printing via direct ink writing (Fig. 1A). The feedstock ink is printed into mechanically stable grid-like architectures that provide the open space and the nutrients required for the mycelium to grow. Incubation of the printed object at ambient temperature and high relative humidity allows for the growth of the mycelium both within and between the deposited filaments without drying of the hydrogel structure.

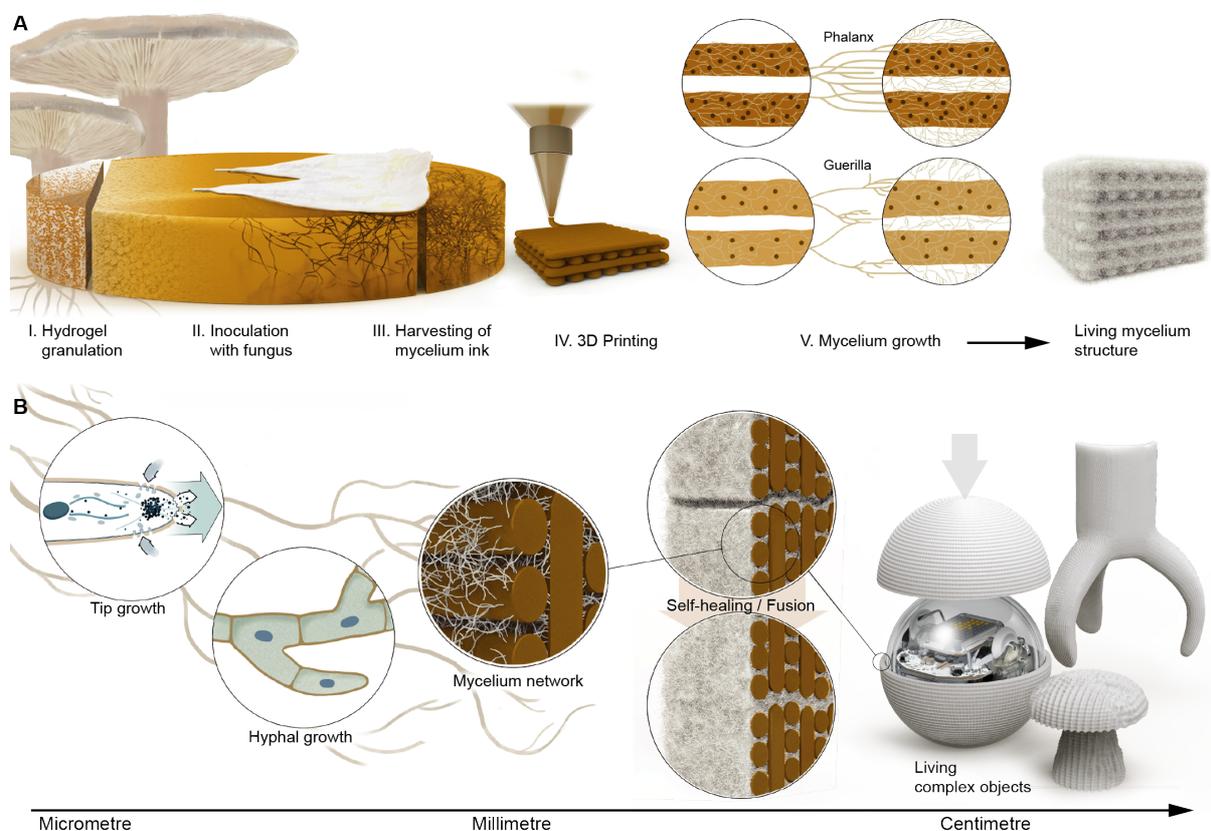

**Fig. 1. Living complex materials and objects made by 3D printing of mycelium-laden hydrogels.** (**A**) (top row) Schematics displaying the distinct steps of the 3D printing process, from the preparation of inks with inoculated mycelia (left) to the direct ink writing of mycelial hydrogels into grid architectures (middle) to the growth of the fungus through *phalanx* or *guerrilla* strategies depending on the nutrient concentration (right). (**B**) (bottom row) Hierarchical structure of the resulting mycelia-based objects, highlighting (from left to right) the cell-level growth through self-organization processes, the hyphae cells that form the mycelial network, the growth of the mycelial network between printed hydrogel filaments, the self-healing and regeneration processes across large air gaps and the macroscopic geometry of the complex material according to the shapes that are relevant for the final applications.

The shaping capabilities of 3D printing can be exploited to manufacture complex materials in geometries that match the functional requirements of specific engineering scenarios. One possible embodiment of this concept is to create protective living skins for robots. To fulfil this function, the skin should be mechanically robust but also sufficiently active to allow for self-regeneration of damaged sites. Such apparently contradicting features can be achieved by 3D printing mycelium-laden hydrogels into engineered structures that would otherwise not exist in the natural world (Fig. 1B). In these structures, mechanical robustness arises from the formation of a strong fibrous mycelial network with a shape that is relevant for the final application. The livingness and the ability to self-regenerate damaged sites emerges from the metabolic activity of the mycelial cells, which have evolved in nature to navigate through and grow within the openings of porous structures. The ultimate function of the living complex material is eventually fulfilled through a unique hierarchical architecture that spans from the individual mycelia cells to the hypha network and grid-like structures shaped into designer macroscopic geometries (Fig. 1B).

To create the proposed living complex material, we first identify the conditions required for the preparation of printable mycelial hydrogels and for fungal growth between the gaps expected in grid-like structures. The processing conditions leading to hydrogel inks with high mycelial concentrations were evaluated by studying the growth behavior of the fungus deposited on top of an agar-based hydrogel substrate (Fig. 2A,B and Fig.S1A,B). To this end, a rectangular piece of fungi inoculum was placed onto the substrate surface and grown at 23 °C and relative humidity of 95% (Supplementary Information). Growth was quantified by measuring the radial and the thickness expansion of the fungus after a fixed period of 5 days upon deposition on hydrogels with distinct initial concentrations of malt extract. The results indicate that mycelial growth is strongly affected by the concentration of malt extract available in the substrate (Fig. 2B). Low concentrations favor radial expansion of the fungus with limited growth of thickness of the mycelium layer on the surface. The radial growth of fungi placed on gels with malt extract contents of 2 - 4 % is around 65 % larger compared to those cultured on top of gels with 15 - 20 % malt extract. Instead of radial expansion, fungi deposited on hydrogels containing these high malt extract contents tend to grow thicker layers on the gel. Indeed, our data reveal that the fungus grows 1.7-times thicker mats on the hydrogel when the initial malt extract concentration changes from 2 to 14 %.

The distinct growth forms observed can be interpreted in terms of the balance between exploration and exploitation used by diverse species of fungi and plants (*17-19*). When exposed to high concentrations of readily available nutrients, mycelia utilize an exploitation growth strategy known as *phalanx*. In this case, the organism advances slowly in a united front, leading to the formation of highly branched and dense hypha mats. In contrast, mycelium growth switches to an exploration form called *guerilla* if the local availability of nutrients is low. Such scenario induces the organism to opportunistically search for nutrients, resulting in far-reaching hyphae with a low level of branching. For the mycelium species utilized in this work, a compromise between exploration and exploitation is found at an intermediate malt extract concentration of 10 %, at which the volume of fungus is approximately two times larger than the levels reached at lower and higher malt contents (Fig. S1C). Because of this balanced exploration-exploitation strategy, mycelium cultured with this optimum nutrient concentration not only grows locally within the hydrogel but are also able to bridge air gaps of up to 2.5 mm at an average growth velocity of 0.20 - 0.35 mm/day (Fig. 2C,D).

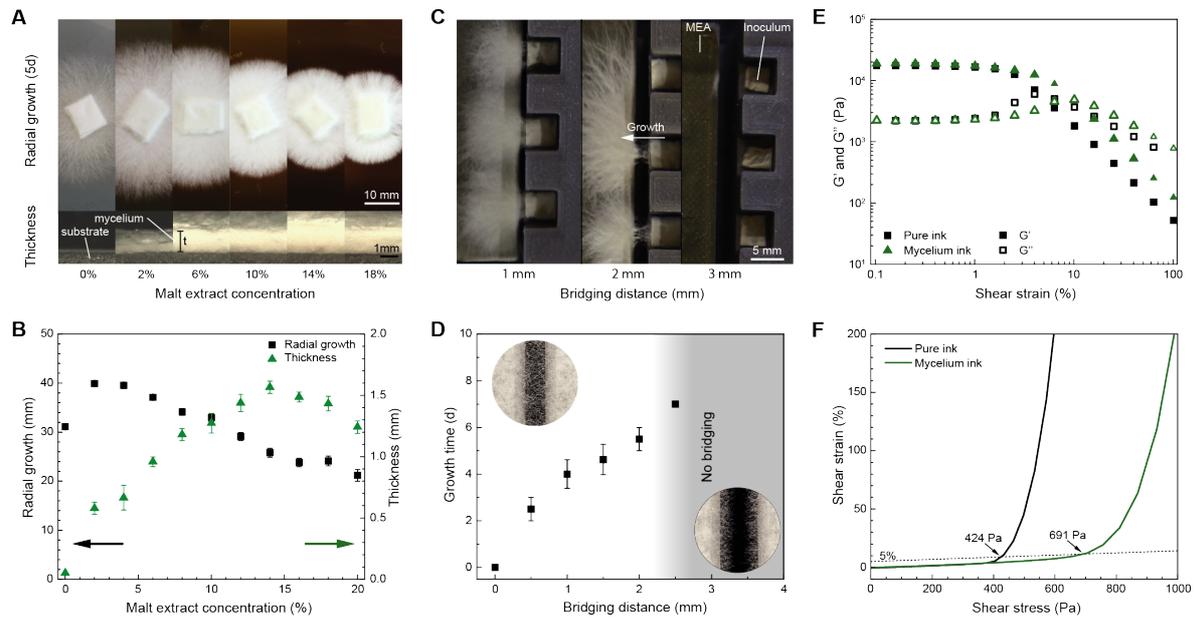

**Fig. 2. Fungal growth behavior and rheology of mycelia-laden hydrogels.** (**A,B**) Radial and thickness growth of mycelia inoculated on the surface of agar gels containing varying concentrations of malt extract after 5 days. The rectangular samples consist of pieces of agar hydrogel with pre-inoculated mycelium. (**C,D**) Bridging of mycelia across air gaps intentionally formed between the surface of inoculated agar gels. The time that it takes for the mycelia to bridge an air gap of defined length is shown in plot (D). Error bars represent the standard deviation of the measurement. (**E**) Storage (G') and loss (G'') moduli of mycelia-laden hydrogel ink as a function of the applied oscillatory strain. (**F**) Flow behavior of the mycelia-laden hydrogel ink, highlighting the yield stress below which no significant deformation takes place. The rheological behavior of a mycelia-free ink is also shown in (E) and (F) for sake of comparison.

3D printing of complex-shaped grids is possible by turning fungal growth media of intermediate malt extract content (10 %) into aqueous inks displaying rheological properties suitable for direct ink writing. For grid-like structures, the printing inks should display a sufficiently high storage modulus to prevent filament sagging and a high yield stress to minimize capillarity- and gravity-driven shape distortions. Using oscillatory rheology, we found that the inoculation of the base ink with mycelium does not alter its viscoelastic properties (G' and G'') at shear strains below 1% (Fig. 2E). However, shear stress-strain measurements reveal that the presence of mycelium increases the yield stress of the base ink from around 400 Pa to nearly 700 Pa (Fig. 2F). Our rheological characterization suggests that the viscoelastic properties at low shear strains are dominated by the constituents of the base hydrogel, whereas the mycelium network plays a mechanical strengthening role that is only activated at higher deformations. Importantly, the storage modulus and the yield stress of the mycelium inks are sufficient to minimize filament sagging and distortions resulting from capillary or gravitational forces (Fig. S2 and supplementary text).

Mycelium-laden inks with optimal rheological properties were 3D printed into stable grid-like architectures using a desktop extrusion-based printer. Grids with distinct filament spacings, diameters and malt extract concentrations were prepared using different ink formulations and programmable printing paths (Fig. S3). Incubation of these printed grids for up to 20 days allowed for the effective growth of mycelium between the hydrogel filaments, leading to mechanically robust living structures (Fig. 3A,B and Fig. S1). The growth of the mycelium over time was quantified by measuring the relative amount of dry biomass left after grinding and extensively washing the grid in water. The results show that the dry mass of mycelium is initially zero and grows linearly up to about 4-5 wt% within the first 10 days of incubation, after which the amount of biomass was found to level off (Fig. 3C). This behavior indicates that the malt concentration of 10 % used in these experiments is sufficient to induce steady mycelial growth in the first 10 days, but the nutrients are eventually depleted from the culture medium after this time window.

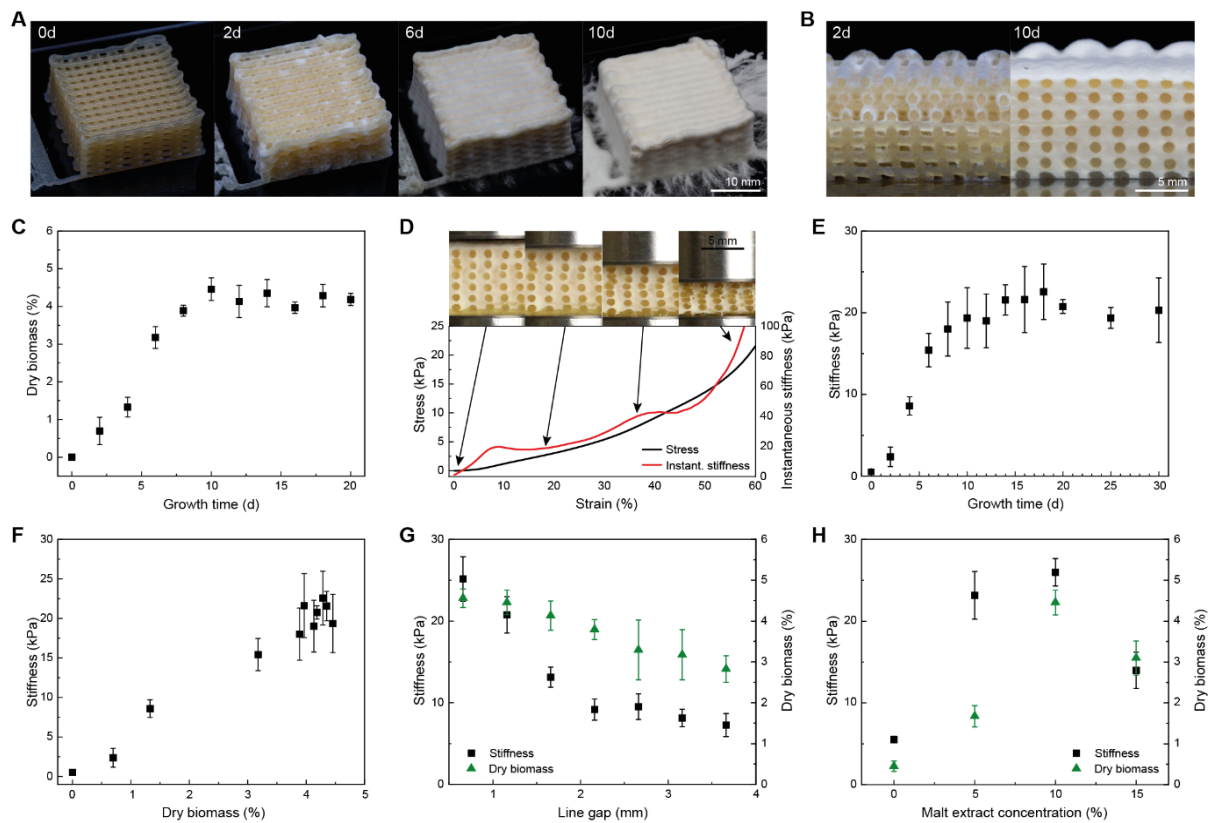

**Fig. 3. Growth and mechanical stiffness of mycelia-based living material with 10% malt concentration.** (**A**) Photographs of the 3D printed mycelial-based grids with increasing growth times. (**B**) Cross-section of the grid after 2 and 10 days of incubation, indicating the significant growth of mycelia in between the printed hydrogel filaments. (**C**) Amount of biomass generated by the fungus as a function of time for grids printed with line spacing of 2 mm and using a nozzle diameter of 0.84 mm (shown in (A) and (B)). (**D**) Representative stress-strain curve obtained from the mechanical compression tests (black line) and the associated variation in instantaneous stiffness (red line). The insets show *in-situ* images of the cross-section of the material at different applied strain levels. (**E**) Apparent stiffness as a function of growth time for mycelia-containing grids printed with line spacing of 2 mm and using a nozzle diameter of 0.84 mm (shown in (A) and (B)). (**F**) Correlation between the apparent stiffness and the biomass concentration of grids shown in (A,B). (**G,H**) Apparent stiffness and biomass content of mycelia-laden grids printed (G) with different gaps between grid lines and (H) with hydrogels containing distinct malt extract concentrations. The apparent stiffness values reported in (E-H) correspond to the instantaneous elastic modulus at a strain value of 25%. Error bars represent the standard deviation of the measurements.

The growth of mycelium between the filaments results in a robust living material that is highly resistant to tearing, stretching and compression compared to the initial hydrogel-based grid (Movie S1). We quantify the stiffness of the grid by performing compression experiments on specimens incubated for different time periods (Fig. 3). Typical stress-strain curves obtained from these measurements reveal a continuous stiffening of the grid as the applied strain is increased. Taking the slope of the stress-strain data as the instantaneous elastic modulus of the material (Fig. 3D), we found that the grid undergoes sequential stiffening processes upon uniaxial compression. In-situ imaging indicates that adjacent filaments along the height of the sample remain separated by the mycelia during the first stiffening process but are eventually pushed into one another when the second stiffening event takes place. As compression and densification proceed, the grid continues to stiffen sharply until it eventually fractures, which is accompanied by a sudden burst of liquid from the sample (Fig. S4 and Movie S1).

To better understand the role of the mycelial network on the mechanical properties of the living material, we measured the elastic modulus of the grid as a function of growth time (Fig. 3E). With a steady increase in the first ten days of incubation, the stiffness of the grid directly reflects the growth curve of the mycelial network. Indeed, the measured stiffness was observed to scale directly with the amount of dry biomass produced by the microorganism (Fig. 3F). This suggests that the elastic modulus of the complex material is determined by the load-bearing capacity of the mycelial network formed between the filaments of the grid. The direct correlation between the stiffness of the grid and the dry biomass produced also helps explain the effects of the malt extract concentration and of the line gap on the elastic modulus of the grid (Fig. 3G,H). For grids prepared with different malt concentrations, we found that an excess of nutrients does not necessarily lead to a higher concentration of biomass within the grid. This might be correlated with the fact that high sugar contents reduce the water activity (*26*), which has been shown to decrease the growth of filamentous fungi (*27*).

The metabolic activity of the mycelial network endows our 3D printed objects with remarkable living and self-regenerating properties. We illustrate these properties by first studying the growth behavior of the microorganism across physically separated surfaces and air gaps in model experiments (Fig. 4A-C and Movie S2). To gain insight into the livingness of our material, the mycelial network that grows between two fungi-laden hydrogel filaments was imaged over time in a laser scanning confocal microscope (Fig. 4B). In this setup, two ink filaments were printed 1.2 mm apart onto a glass slide and covered with another thin glass slide leaving an air gap in between. The mycelial network that grows from the surface of the filament was detected using a fluorescent dye (calcofluor white) that specifically adsorbs onto the chitin molecules present in the walls of the hypha cells.

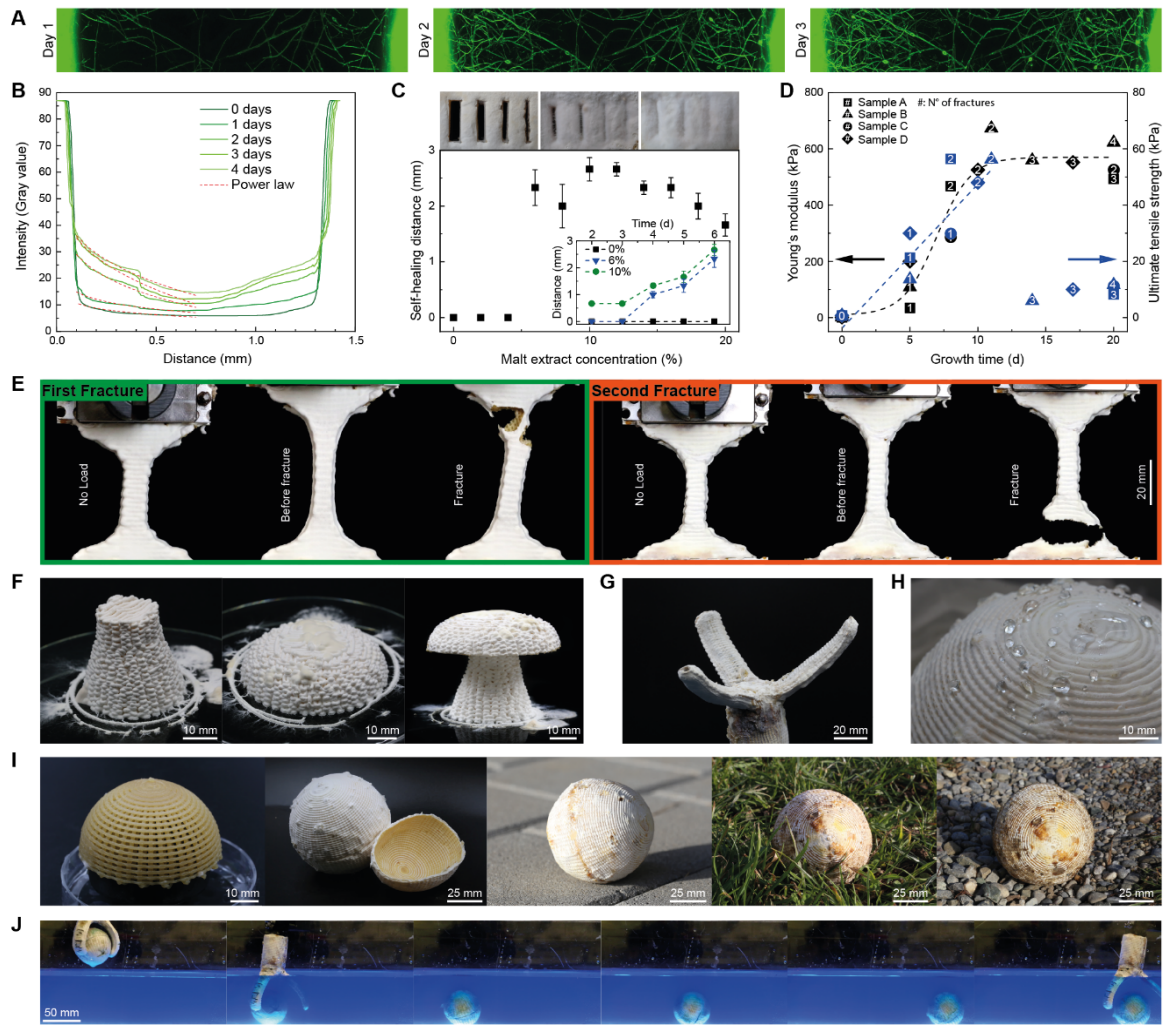

**Fig. 4. Livingness, self-healing and application of mycelia-based living materials.** (**A**) Laser confocal microscopy images of dyed mycelia (green) growing between two hydrogel filaments printed 1.2 mm apart. (**B**) Intensity profiles of the images shown in (A) quantifying the time evolution of the spatial distribution of the mycelial network grown between the two filaments. The decay in fluorescence intensity as a function of distance is fitted with a power law to demonstrate the scale-free nature of the network. (**C**) Self-healing distance covered by the mycelia across a cut deliberately introduced into a mycelium-containing film prepared with different malt extract concentrations. The pictures shown as insets display a film with 10 % malt extract after 2, 4 and 6 growth days. The distance healed by the mycelium over time is also included as an inset for specimens with selected malt extract concentrations. (**D**) Young's modulus and ultimate strength of mycelia-laden samples subjected to multiple fracture-healing cycles during growth of the living material. (**E**) Photographs of the tensile mechanical tests performed to demonstrate the self-healing ability of the mycelia-based material. (**F**) Joining of 3D printed parts into a single object through growth of mycelia between surfaces in physical contact. (**G,H**) Living skins printed on top of (F) a robotic gripper and (G) an untethered rolling robot. Water droplets on the surface of the skin illustrate the hydrophobic nature of the living material. (**I**) Snapshots of the skin covering the rolling robot before and after mycelial growth. Photographs of the rolling robot after friction-driven locomotion on different surfaces demonstrate the mechanical robustness of the living skin. (**J**) Snapshots showing the underwater cooperative interaction between the robotic gripper and the untethered rolling robot protected by the living skin. Error bars represent the standard deviation of the measurement.

Confocal images of the mycelium formed in the air gap between the printed filaments reveal that the microorganism forms a fractal-like network during growth under the conditions of the experiment. Such fractal morphology is typically characterized by a decrease in the number of cells ($N$) with the distance from the filament surface ($L$), as expressed by the scaling relation: $N \sim L^D$, where $D$ is the fractal dimension. By assuming the fluorescence intensity to be an indirect measure of the cell number and fitting this equation to our experimental intensity data, we estimate a fractal dimension that lies between 1.16 and 1.32 for the mycelial network grown for 2 - 4 days. These values fall within the typical range of mycelium networks, which may vary from 1 for organisms that grow by the *guerrilla* strategy to 2 for fungi with *phalanx* growth form (*28*). The fact that the fractal dimension obtained from our image analysis lies closer to 1 reflects the strong *guerrilla* growth characteristics and exploratory nature of our mycelial network.

The exploratory nature of the mycelial network is key to enable the regeneration of damaged areas of the living material caused by its interactions with the environment. To demonstrate this self-regenerating capability, we introduced cuts of different sizes in a piece of living mycelial film and measured the distance travelled by the growing front as a function of time for specimens prepared with different malt extract concentrations (Fig. 4C). The results show that the mycelium grows at a speed in the range 0.6 – 0.7 mm/day after an initial lag phase until a maximum healing distance of 2.5 to 3 mm, if the malt concentration is equal or higher than the threshold value of 6 %. This size range is comparable to the defect size of 5 mm typically used to study self-healing processes that take place during regeneration of fractured human bone (*29*). While the availability of nutrients is crucial to enable self-regeneration, viability experiments show that the mycelial network can be brought back to life even after the material has been depleted from culture medium in the hydrated or dried state for as many as 8 months (Fig. S11). Moreover, the fungi generated after this dormant phase is able to outcompete other microorganisms, such as yeast, during the re-growing phase. This impressive resilience preserves the livingness of the mycelia network even in very adverse environments and conditions.

To evaluate whether such self-healing effect also enables recovery of the mechanical properties of the living material, we measured the tensile strength and elastic modulus of dogbone-shaped specimens before and after successive fracture-healing events (Fig. 4D and Movie S3). Considering the living nature of the material, the experiments were carried out during the growth phase of the mycelial network. Remarkably, the fractured complex material is able to self-heal into stronger and stiffer structures, as long as the mycelial network has sufficient nutrients to grow across the fractured surfaces. In fact, the network formed across fractured surfaces was observed to be strong enough to shift the fracture site to another location of the sample in a subsequent healing cycle (Fig. 4E). After the initial growth phase, the healed material still maintains a high elastic modulus but fails at much lower stresses. In addition to self-healing damaged sites of a given object, we found that the mycelium is also able to grow across the air gaps between separate objects positioned next to one another (Fig. S6A). This enables the creation of complex-shaped living structures by simply joining individual parts that were manufactured separately (Fig. 4F).

The livingness and printability of the mycelium-based hydrogel opens the possibility to create functional structures with bespoke designs and unprecedented adaptive behavior. As illustrative examples, we 3D printed our living material in the form of self-regenerating functional skins for a robotic gripper and for an untethered rolling robot (Fig. 4G-I). Following the strategy described earlier, the skins are manufactured by printing the mycelium-laden hydrogel into a grid-like architecture that conforms to the geometry of the gripper's arms and of the spherical shape of the rolling robot (Fig. S7). After growth, the skin is sufficiently robust to keep its mechanical integrity and living nature while exposed to the high frictional forces required to move the rolling robot on a variety of different surfaces (Fig. 4I and Movie S4). Moreover, the hydrophobic character of the mycelial network prevents wetting of the skin by water, providing water-proof and Lotus leaf-like self-cleaning capabilities to the rolling robot. We demonstrate this additional feature by designing an experiment in which the skin-covered gripper and rolling robot work cooperatively to enable programmed underwater motion (Fig. 4J and Movie S5). The mechanical robustness, softness, self-healing capability and water-proof nature of the living material provide robots with a protective skin that feature several functionalities of biological animal skins.

In summary, combining the livingness of microorganisms with the shaping capabilities of 3D printing technologies is a powerful pathway towards the creation of functional living materials with unparalleled complex adaptive properties. Using this strategy, hydrogels loaded with the microorganisms of interest can be shaped into architectures that fulfil a functional design and also provide an adequate environment for growth of the biological species. The living properties of the resultant architectures emerge from the metabolic activity of the organisms embedded in the hydrogel. This metabolic activity imbues the living material with several hallmarks of complex adaptive systems, including the dissipative self-organization processes that enable growth and regeneration, the hierarchical organization of building blocks across multiple length scales, the optimal transport properties of scale-free fractal networks and the decision-making capabilities that emerge from the decentralized cooperative action of information-processing cells. The three-dimensional printing of mycelium hydrogels with such complex adaptive properties offers a unique opportunity to create functional living materials for several applications and can potentially inspire other strategies to bring life to the realm of materials.

**Acknowledgements:**

**Funding:**

Swiss Competence Center for Energy Research (SCCER) - Capacity Area A3: Minimization of energy demand

Swiss National Science Foundation Consolidator grant BSCGIO_157696

Swiss National Science Foundation within the framework of the National Center of Competence in Research for Bio-Inspired Materials.

**Author contributions**

Conceptualization: SG, PR, KM, ARS

Methodology: SG, EC, JK, ET, FC, KM, PR

Software: SG, FC

Investigation: SG, EC, JK, ET, FC, KM, PR

Visualization: SG, EC, JK

Funding acquisition: ARS

Project administration: SG

Supervision: SG, KM, PR, ARS

Writing – original draft: SG, ARS

# Supplementary Information

## Three-dimensional Printing of Mycelium Hydrogels into Living Complex Materials

Silvan Gantenbein, Emanuele Colucci, Julian Käch, Etienne Trachsel, Fergal B. Coulter, Patrick A. Rühs, Kunal Masania1*, André R. Studart *

Correspondence to: k.masania@tudelft.nl, andre.studart@mat.ethz.ch

**This PDF file includes:**

    Materials and Methods
    Supplementary Text
    Figs. S1 to S10
    Captions for Movies S1 to S5

**Other Supplementary Information for this manuscript include the following:**

    Movies S1 to S5

**Materials and Methods**

Materials

The following chemicals were purchased and used as received: malt extract (Sigma-Aldrich), peptone (Sigma-Aldrich), agar (Duchefa), κ-carrageenan (Acros Organics), maltodextrin (PanReac AppliChem), calcium sulphate ($CaSO_4$, Sigma-Aldrich), Sylgard 184 (Sigma-Aldrich), calcofluor white (Sigma-Aldrich). A cellulose-based thickener (MCG, Vivapur MCG 811P) was kindly provided by JRS Pharma. The fungal species *Ganoderma lucidum* (China strain number 112001) and beech wood sawdust were purchased from gluckspilze.com

Preparation of growth media and cell culture

Two types of growth media were prepared as feeding substrates for the fungus: malt extract agar (MEA) and wood chip substrate. Malt extract agar medium was prepared by adding 0 – 200 g/L (0 – 20 %) malt extract, 5 g/L (0.5 %) peptone and 16 g/L (1.6 %) agar to water and adjusting the pH to 6.8 using 1 M NaOH and 1M HCl. To obtain MEA plates, the medium was sterilized at 121 °C for 20 min in an autoclave (2840 EL, Tuttnauer) and, subsequently, poured into cell-culture dishes and let to solidify. The wood chip-based substrate was prepared by mixing 800 g beech wood sawdust with 8 g maltodextrin, 10 g calcium sulphate and 1 L distilled water. The mixture was placed into glass jars and autoclaved at 121 °C for 20 min.

The fungus was subcultured by placing a small slice of a MEA plate fully covered with mycelium onto a fresh growth plate (2 % malt extract) every 3 - 5 weeks. The cultures were grown in the dark at 23 °C and 95 % relative humidity.

Growth behavior on agar plates

The growth of the fungus on agar-based substrates was evaluated by preparing MEA plates containing malt extract concentrations ranging from 0 to 20%. After placing an inoculum (1x1 cm) into the middle of the culture plate, samples were stored in the dark at 23 °C under 95 % relative humidity and the radial expansion of the fungus was measured every day. The thickness of the mycelium layer was determined optically by cutting out a piece of the plate and observing its cross-section under an optical digital microscope (VHX-6000, Keyence).

Bridging of mycelium across air gaps

The ability of the mycelium to span gaps over air was determined by placing inoculated MEA plates at a well-defined distance from each other using customized polylactic acid sample holders made in a desktop 3D printer (Ultimaker 3). The sample holder contains two ledges separated by a deep gap to prevent the mycelium from simply growing on the surface between the MEA plates (Fig. S8). For the experiment, a piece of overgrown MEA substrate is placed on the ledge opposite to a fresh substrate such as to create air gaps ranging from 0.5 mm to 6 mm. The samples were stored at 23 °C and 95 % relative humidity and monitored every day to evaluate the growth of the mycelium across the air gap.

Preparation of mycelium inks

Mycelium inks for 3D printing were prepared using the same medium compositions utilized for the preparation of the MEA plates. Such base growth medium consists of an aqueous solution containing 0 – 20 % malt extract and 0.5 % peptone at a pH of 6.8. To adjust the rheological properties of the base medium, the ink also contains 15 g/L (1.5 %) agar, 15 g/L (1.5 %) κ-carrageenan and 30 g/L (3 %) of cellulose-based thickener (MCG) pre-suspended in water. The suspension of cellulose-based thickener was prepared by dispersing the MCG powder in deionized water using a high-shear mixer (T25 digital Ultra Turrax, IKA) for 5 min at 1,000 rpm followed by another 5 min at 20,000 rpm. After a rest period of at least 15 minutes, the suspension was combined with the remaining ink components and autoclaved at 121 °C for 20 min. The sterilized and solidified hydrogel was fragmented into a microgranular gel by mechanical mixing using a dynamic mixer (A-302, AIMOSHI) equipped with a milk frother tip (SM 3590, Severin) at 5,000 rpm for 1 min.

The ink was either used directly as a reference ink or inoculated with mycelium to create mycelium-containing inks. The mycelium inks were inoculated by spreading the fresh ink in a cell culture dish and placing a fungi inoculum onto the ink surface. After 7-10 days at 23 °C and 95 % relative humidity, the whole surface was covered with mycelium and the dense mycelium layer formed was removed. The inoculated ink remaining underneath was quickly mixed by hand and filled into the printing cartridge. Before printing, the cartridge was centrifugated (Z306, Hermle) at 2,000 rpm for 2 min to remove any entrapped air.

Rheology of mycelium inks

The rheological behavior of the inks before and after inoculation was characterized using a plate-plate geometry mounted on a strain- and stress-controlled rheometer (MCR 302, AntonPaar). Oscillatory shear tests were performed to measure the storage (G') and loss (G'') moduli of the inks by applying amplitude sweeps at an angular frequency of 1 rad/s and frequency sweeps at 1 % shear strain. The yield stress of the inks was determined by applying a stress-controlled deformation to the sample and measuring the resulting shear strain. The yield stress was arbitrarily taken as the stress required to achieve a plastic deformation of 5 %. The elastic recovery of the inks was measured by applying a shearing protocol designed to partly simulate the printing conditions. In this protocol, the storage and loss moduli of the undisturbed ink is first probed by applying an oscillatory shear strain of 1 % amplitude at 10 rad/s. Afterwards, the material is subjected to steady shear at 1 s-1 and the apparent viscosity is measured. The elastic recovery of the ink is finally quantified by measuring the evolution of the storage and loss moduli of the ink through oscillatory measurements right after the steady shear was ceased. In addition to elastic recovery, the inks were also characterized in terms of flow behavior under steady shear conditions. To this end, we measured the apparent viscosity of inks containing varying concentrations of malt extract and rheological modifiers at shear rates between 0.1 and 100 s-1.

3D Printing of mycelium inks

Grid structures were 3D printed using a desktop printer (Ultimaker 2+) modified to enable the deposition of filaments via direct ink writing. For this purpose, the original print head was replaced by a custom-made extrusion system consisting of a mechanically driven syringe pump that can accommodate 25 ml syringes (Fig. S9). The print paths for the 3D printed grid structures were designed using Grasshopper (Rhinoceros, Robert McNeel & Associates). The custom code enabled the 3D printing of grids with various sizes, nozzle diameters and print line distances. Typically, parts were printed onto sterilized glass substrates using a printhead velocity of 20 mm/s and the printer was operated inside a laminar flow hood. Unless otherwise stated, the grid samples were printed with a 0.84 mm nozzle and a line gap of 1.2 mm and were later grown for 10 days at 23 °C at 95 % relative humidity. The printed structures were found to remain dimensionally stable under this high humidity (no swelling). Pictures of the grids after mycelial growth were taken using a Canon EOS 6D and a macro objective (EF 100mm f/2.8L Macro IS USM, Canon). Cross-section images were obtained by cutting the sample in half using a razor blade.

Biomass of mycelia-containing grids

The biomass of printed grids containing mycelium was determined by extracting the material generated by the fungi during growth. To this end, the samples were first cut into small pieces and immersed in water. After shaking the samples in water for 5 minutes, the resulting suspension was filtered using a 100 µm woven wire mesh (stainless steel, Retsch) and rinsed with water to wash away any remaining ink components. The solid residue remaining in the mesh was weighed and dried at 60 °C for 24 h before weighing again. The dry biomass was determined as the fraction of the dried mass with respect to the initial weight of the grid.

Mechanical Compression tests

The mechanical properties of 3D printed grid structures with nominal size of 20x20x10 mm were measured by performing uniaxial compression tests. The compression tests were carried out using a universal mechanical testing machine (AGS-X, Shimadzu) with a 100 N capacity load cell by applying a constant displacement rate of 5 mm/min. The mechanical stiffness of the samples was taken at an arbitrarily applied strain between 24 and 26 %. With the help of cyclic measurements, we found that the material shows reversible viscoelastic properties when compressed at such a strain value. The instantaneous stiffness data shown in the main text (Figure 3) shows the same trends irrespective of the arbitrarily chosen strain value.

Confocal microscopy of growing mycelium

Scanning laser confocal imaging was performed to study the growth of mycelium between two printed filaments separated by an air gap (Fig 4A). The filaments were 3D printed using a mycelium ink containing 10 µg/g of calcofluor white to enable staining of the chitin molecules produced by the fungus. The ink was printed directly onto a Nr. 0 glass slide with a thickness of 130 µm and covered with another glass slide using PDMS frames as spacers (Sylgard 184). The PDMS frame acted as a vapor barrier to prevent drying of the samples, while also allowing for the diffusion of oxygen needed for cell growth. The samples were imaged using a scanning laser confocal microscope (TCS SP8, Leica) at 20x magnification (HC PL Fluotar 20x/0.55, Leica). Fluorescence was induced using a 405 nm diode laser and then collected with a hybrid detector for wavelengths from 439 nm to 713 nm. For the time-lapse experiments, images with 512 x 2812 x 80 pixels and a voxel size of 0.506 x 0.506 x 1.275 µm were captured every 25 minutes. Images are displayed as maximum z-projections and image analysis was performed using Fiji image analysis software (30).

Self-healing of mycelia-based living materials

The self-healing behaviour of the living materials were first studied by performing tensile mechanical tests on mycelia-containing grids printed in a dogbone geometry (Fig. 4C,D) The grids were printed with a nozzle diameter of 0.84 mm, a line gap of 1.2 mm and a malt extract concentration of 10 % in the ink. The dogbone samples showed a transverse cross section of 10 x 10 mm, a gauge length of 36 mm and an overall length of 60 mm between end tabs. The specimens were fixed by end tabs consisting of two aluminum meshes covering a core of inoculated wood chip substrate and spaced apart by metal nuts. The two metal end tabs were fixed together using aluminum bars to prevent the samples from deforming during mounting in the testing machine (Fig. S6B). The metal bars were removed prior to testing and the samples were tested at a displacement rate of 10 mm/min using a universal mechanical testing machine with a 100N capacity load cell (AGS-X, Shimadzu). After fracture, the sample was returned to the initial distance and the aluminum bars reattached to enable self-healing of the fractured surfaces. Self-healing occurred by incubating the samples at 23 °C and relative humidity of 95 % for at least 3 days. Following the incubation period, the self-healed samples were tested again using the procedure described above. This procedure was repeated multiple times to characterize the mechanical properties of the sample after several fracture-healing events (Fig. 4C).

The self-healing ability of mycelium-covered films was investigated by deliberately cutting defects with increasing width into overgrown MEA plates (Fig. 4B). The films were prepared with malt extract concentrations varying from 0 to 20 %. The self-healing ability of the film was characterized in terms of the length of mycelium that grew across the cut width.

Living skins for robotic gripper and rolling robot

A custom-built 5-axis 3D printer (Stepcraft D-600 Gantry with ISEL ZDS 2030 Tilt Rotary Table) was equipped with a volumetric dispensing unit (eco-PEN 300, ViscoTec) (31) to create the living skins for the robotic gripper and the untethered rolling robots (Fig. 4F-I). The grid-like architecture used as template for the skin was printed upon non-planar mandrels that were fabricated from polylactic acid using a regular desktop 3D printer (Prusa mk3, Prusa). Multi-axis CNC toolpaths were calculated, and the subsequent Machine GCodes were created in the visual programming environment Grasshopper3D. Such toolpath strategies were essential to ensure that the printer nozzle was maintained perpendicular to the surface of the substrate at all times during the course of the print, enabling a smooth and even coating. The living skins were formed after incubation of the as-printed grids at 23 °C and 95 % relative humidity for 7 days. The robotic gripper comprised a custom-made three-finger gripper mounted on a commercial robotic arm (X-arm 2.2, Lobot), whereas the rolling robots consisted of commercially available products in two distinct sizes (Bolt & Mini, Sphero). The living skin of the rolling robots was generated by covering the spherical robot with two mycelial grids with hemisphere shape (Fig. 4H) and joining them together through the air bridging of mycelia across opposite surfaces. Similarly, the skin for the fingers of the robotic gripper was assembled from half-shells, whereas flat grids were wrapped around the base of the robot and left to join for 3 days to form a continuous skin.

Contact angle measurements

The hydrophobic nature of the living skin was characterized by measuring the contact angle of water on the surface of a MEA plate covered with mycelium (Fig. S10). Contact angle measurements were conducted using a drop shape analyzer (DSA 100, Krüss). The analyzed water droplets had a volume of 8 µL and were deposited at a dispensing rate of 20 µL/min using a 0.51 mm nozzle.

Scanning electron microscopy

Cross-sections of mycelium covered grids were freeze-dried (FreeZone 2.5, Labconco) and imaged with a scanning electron microscope (Gemini 450, Zeiss) operated with an acceleration voltage of 3 kV. Prior to imaging, the samples were coated with a 5 nm Pt layer using a compact coating unit (CCU-010, safematic).

**Supplementary Text**

Gravitational forces on printed grid structures

3D printed grid structures are prone to distortion due to capillary and gravitational forces. Previous work has shown that a minimum elastic modulus is needed to prevent excessive sagging of suspending filaments in a printed grid (32, 33). We have experimentally observed that gravity can also reduce the overall height of the grids if the ink is not sufficiently strong to withstand the local compressive stresses developed at the intersections between orthogonal filaments.

A simple model based only on gravitational forces was developed to estimate the yield stress required to print stable grids with minimum height reduction. The model assumes that the highest stresses in a self-supporting grid occurs in the base layer, as a result of the mass of printed ink that is stacked upon it. For the hypothetical case of an infinite grid, it is convenient to represent the structure by a periodic unit cell that contains all the weight that is expected to be carried by the base layer (Fig. S2E). Such a unit cell consists of two orthogonally arranged filaments with diameter $D_N$ and length $L_D$. To form the three-dimensional structure, the unit cell is repeated periodically within the plane and is stacked vertically to create a grid of defined height, h. The weight of this unit cell on the base layer defines the stress that the grid needs to withstand to prevent gravity-induced distortion. Considering that the gravitational forces are applied to the ink at the intersection between filaments in the base layer, the resulting maximum stress is given by:

$$\sigma = \frac{\rho V_{tot} g}{A} \qquad (S1)$$

where $\rho$ is the density of the ink, $V_{tot}$ is the total volume of unit cells stacked vertically in the grid, $g$ is the gravitational constant and $A$ is the intersection area formed between two orthogonal filaments.

The total volume $V_{tot}$ can be calculated from the volume of an individual filament of the unit cell ($V_f$):

$$V_{tot} = V_f (N - 1) \qquad (S2)$$

where $N$ is the number of layers needed to make up the whole height of the grid.

The volume $V_f$ depends on the diameter ($D_N$) and length ($L_D$) of the filament (Fig. S2E) in the unit cell, as follows:

$$V_f = \frac{\pi D_N^2 L_D}{4} \qquad (S3)$$

Here, the filament diameter is assumed to be equivalent to the nozzle diameter. The length $L_D$ corresponds to the distance between two print lines and is given by:

$$L_D = L_G + D_N \qquad (S4)$$

where $L_G$ is the line gap used as printing parameter.

The number of layers $N$ can be calculated from the grid height and the nozzle diameter:

$$N = h/D_N \qquad (S5)$$

Assuming the cross-sectional area $A$ to be $D_N^2$ and that $N \gg 1$, we can combine the above equations to obtain the maximum stress $\sigma$:

$$\sigma = \frac{\pi \rho g}{4} \frac{L_D}{D_N} h \qquad (S6)$$

The reduction in grid height resulting from gravitational forces can be prevented if the yield stress of the ink is higher than the maximum stress applied: $\sigma_y > \sigma$ (Fig. S2E).

Structure and mechanics of grids with varying filament diameter

The mechanical properties of grids printed with distinct line gaps and malt extract concentrations were found to be directly correlated with the dry biomass (Fig. 3G,H). In contrast, compression experiments on grids with varying filament diameters and fixed line distance showed that the mechanical stiffness of such structures does not correlate with the dry biomass (Fig. S5). This apparent inconsistency suggests that other factors may influence the mechanical properties of grids printed using distinct nozzle diameters.

    To interpret these results, it is important to note that an increase in the filament diameter through larger nozzles is accompanied by a reduction of the line gap of the grid (Fig. S3B). Since smaller line gaps result in more mycelium (Fig. 3G), one would expect larger filament diameters to lead to grids with higher biomass content. This is not confirmed by our experiment data (Fig. S5), which show that the dry biomass decreases with filament diameter for nozzles larger than 0.6 mm. The observed results indicate that the small gaps in grids printed with larger nozzles might confine mycelium growth and thus lead to a denser load-bearing network in these grids. Such a denser mycelium network is a possible explanation for the increase in mechanical stiffness of grids with larger filaments, in spite of their lower biomass content (Fig. S5). Indeed, the fractal nature of the network is expected to result in overall denser mycelium structures in grids with smaller line gaps, corroborating to the above interpretation.

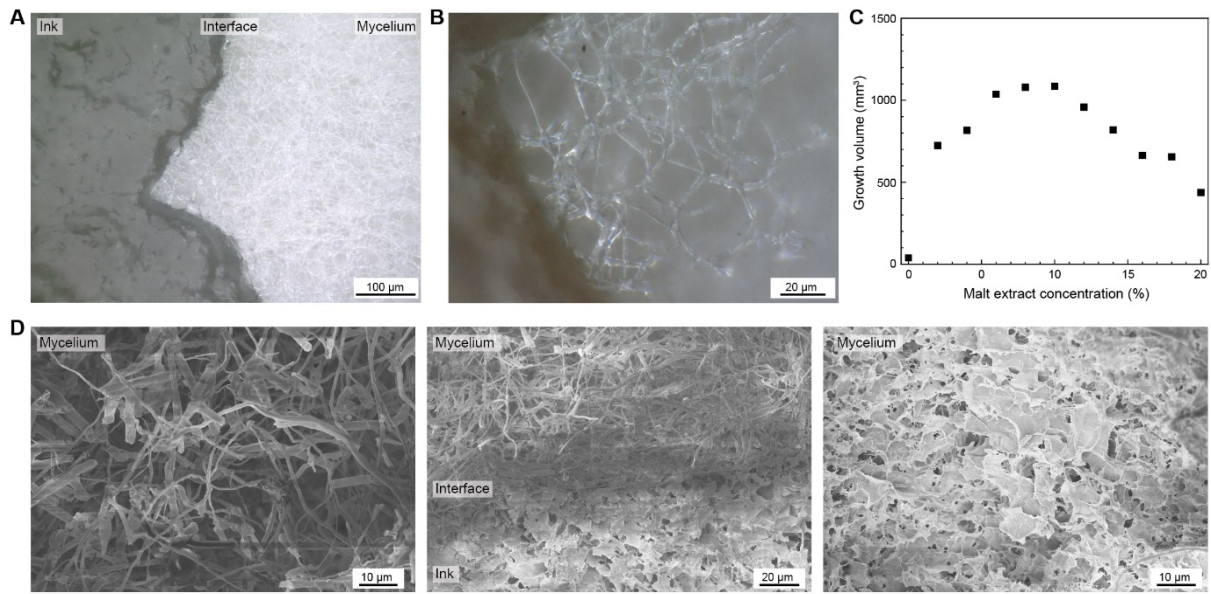

**Fig. S1. Mycelium growth within hydrogel grid**

(**A,B**) Optical microscopy images of a cross-sectional cut of a 3D grid printed from an ink containing 10% malt extract concentration. The images show the mycelium growth on the surface of the ink. (**C**) Estimated volume of the mycelium layer grown on agar hydrogels. The volume was calculated from the radius and thickness of the grown mycelium network (Fig. 2B). (**D**) Scanning electron microscopy images showing the freeze-dried structure of mycelium grown on a printed hydrogel filament.

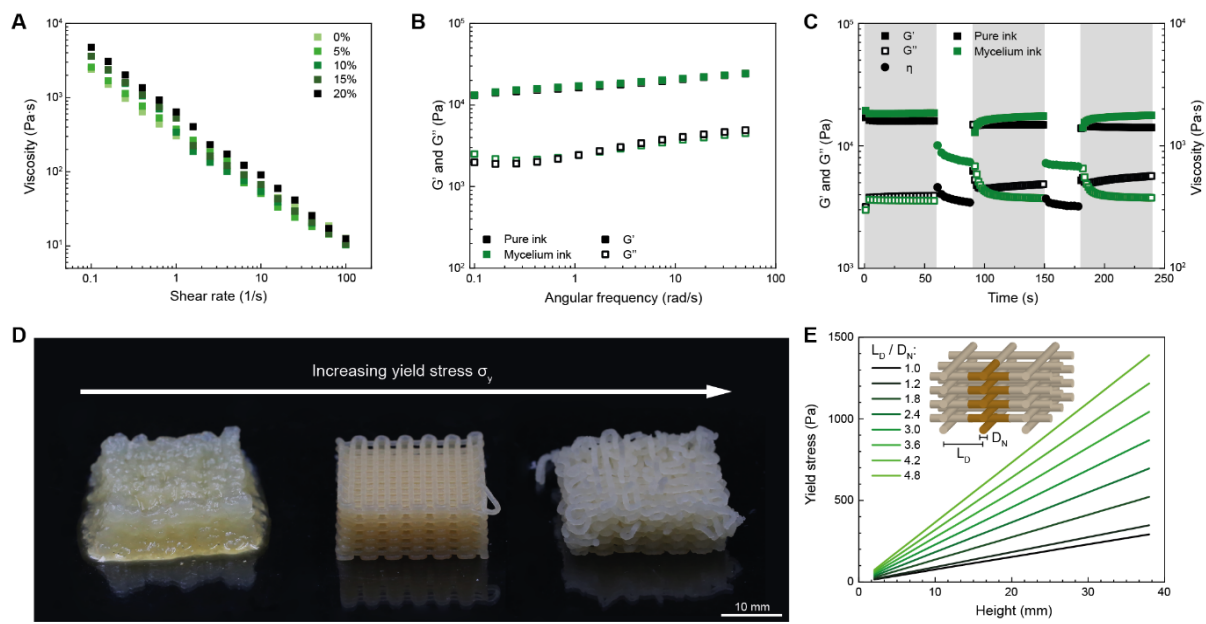

**Fig. S2. Rheological properties and printability of mycelium inks.**
(**A**) Apparent viscosity of hydrogel inks measured under steady shear conditions, indicating the shear thinning behaviour of inks containing varying concentrations of malt extract. (**B**) Storage (G') and loss (G'') moduli of inks before and after inoculation obtained from oscillatory rheological measurements at increasing applied frequency. The data show that the printed inks are predominantly elastic over the entire range of probed frequencies. (**C**) Rheological response of inks under printing simulation conditions. The alternating oscillatory (grey area) and steady-state (white area) measurements show quick recovery of the storage modulus of the ink upon cessation of shear. (**D**) 3D printed grids obtain from inks with increasing yield strength of approximately 100 Pa (left) 400 Pa (middle) and >1000 Pa (right). The images indicate filament sagging and flowing for inks with low yield stress, whereas high-yield-stress inks lead to fragile and brittle grids. Optimal print results are achieved with inks at intermediate yield stress level. (**E**) Effect of height and line gap spacing on the minimum yield stress required to print hydrogel grids that resist gravity-induced distortion. $L_D$ and $D_N$ correspond to the line spacing and the filament diameter, respectively.

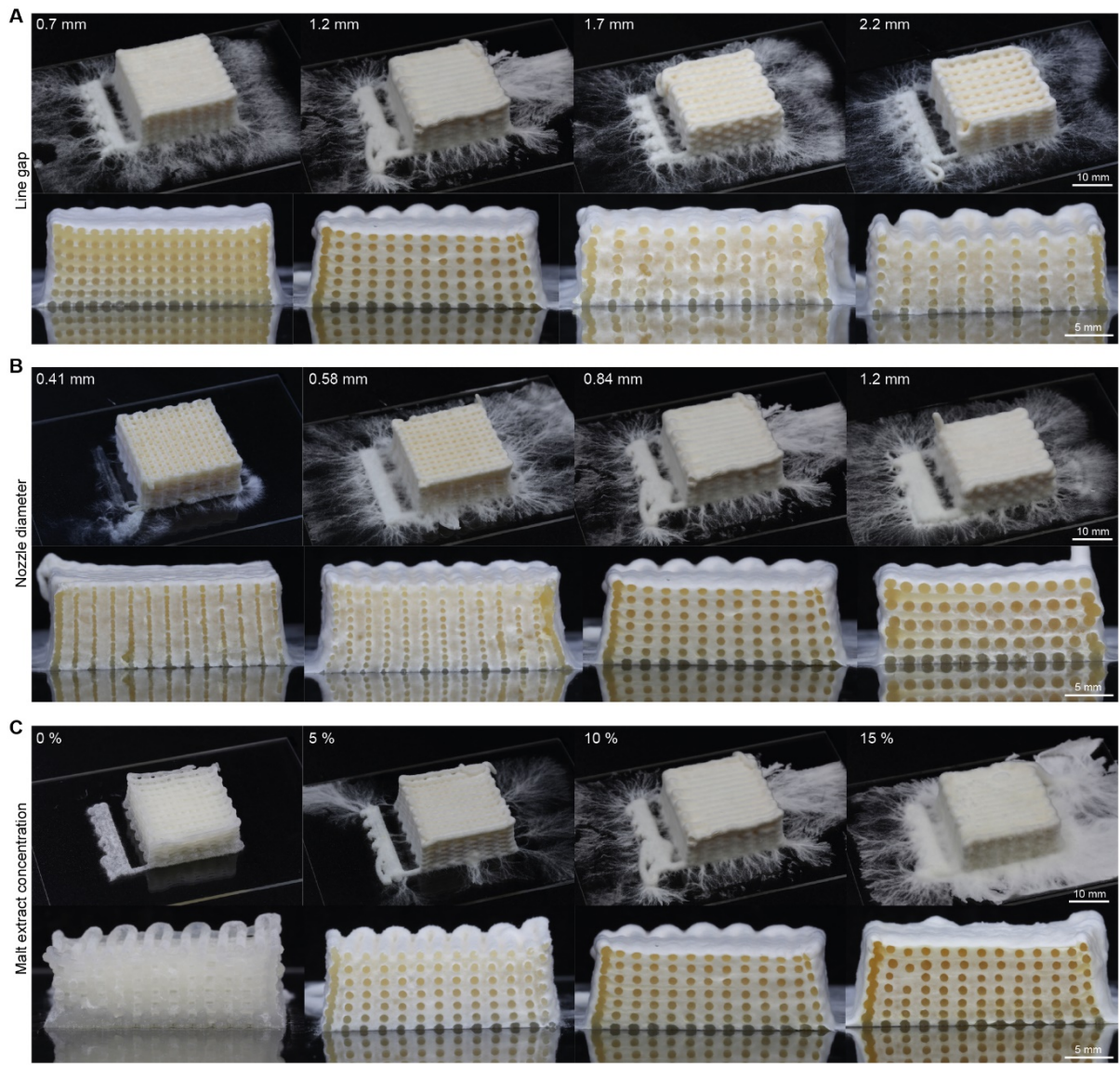

**Fig. S3. 3D printed mycelium grid structures.**
(**A-C**) Photographs of grids after 10 days of growth for structures printed with varying (A) line gaps, (B) nozzle diameters, and (C) malt extract concentrations.

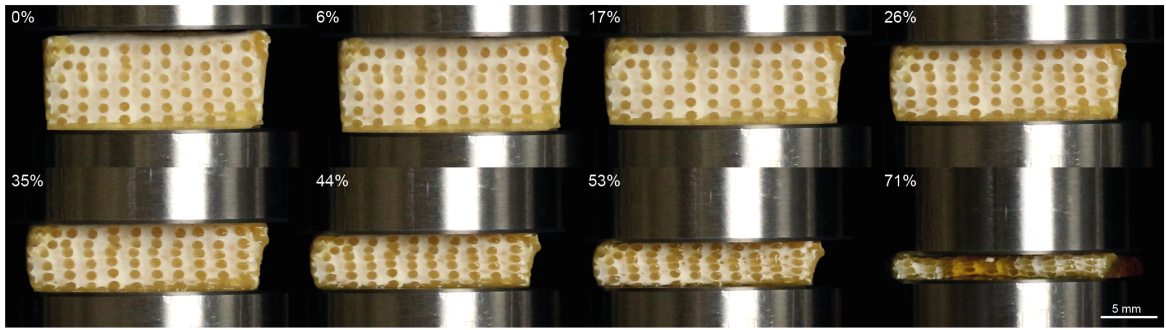

**Fig. S4. Mycelium grid structure during compression testing.**

Snapshots taken at distinct compressive strains show the changing internal structure of the material. The grids were printed from inks containing 10 % malt extract with a line gap of 1.2 mm and nozzle diameter of 0.84 mm.

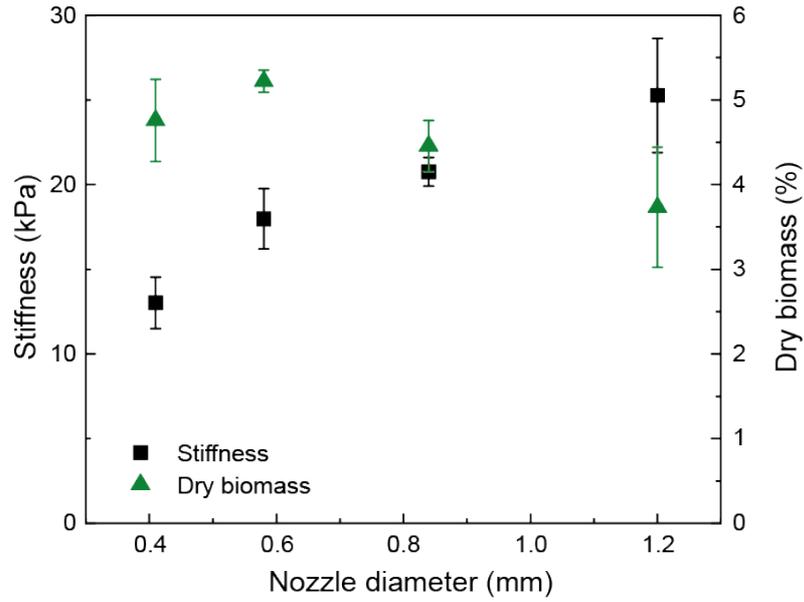

**Fig. S5. Effect of nozzle diameter on the dry biomass and stiffness of mycelium grids.**
Apparent mechanical stiffness and biomass content of mycelia-laden grids printed with varying nozzle diameters. The line distance and the malt extract concentration were kept constant at 2 mm and 10%, respectively. The apparent stiffness values reported correspond to the instantaneous elastic modulus at a strain value of 25%.

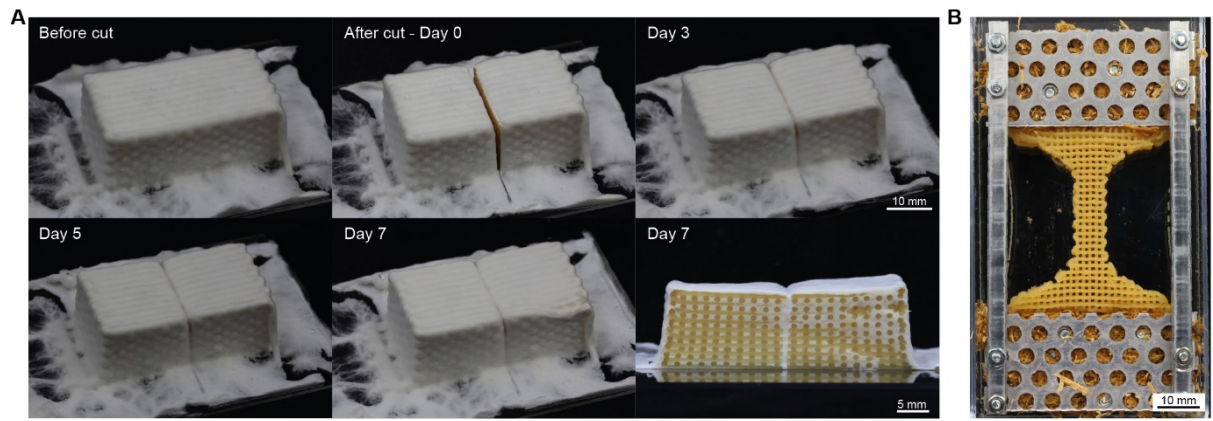

**Fig. S6. Self-healing of cuts made in the printed mycelium grids.**

(**A**) A 3D printed grid was intentionally cut and imaged for a duration of 7 days. After this time period, the cross-section image of the sample shows that the cut is fully healed through the growth of mycelia across the open space. (**B**) Setup comprising metal frame and wood chip substrate used to evaluate the tensile mechanical properties and self-healing ability of printed mycelium grids.

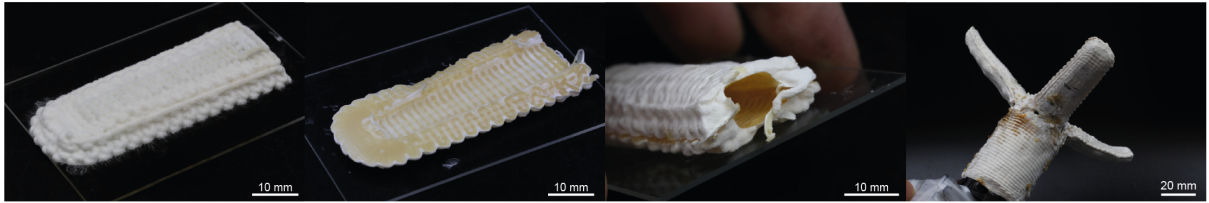

**Fig. S7. Strategy to fabricate the mycelium-based skin for the robotic gripper.**
From left to right: non-planar 3D printed half of the robotic skin after 7 days of growth. Counterpart flipped to show the bottom side of the printed structure. The two halves put together ready to mount onto the mechanical actuators. Final mycelium-based skin assembled around the gripper after an additional 3 days of growth to fuse the individual parts together.

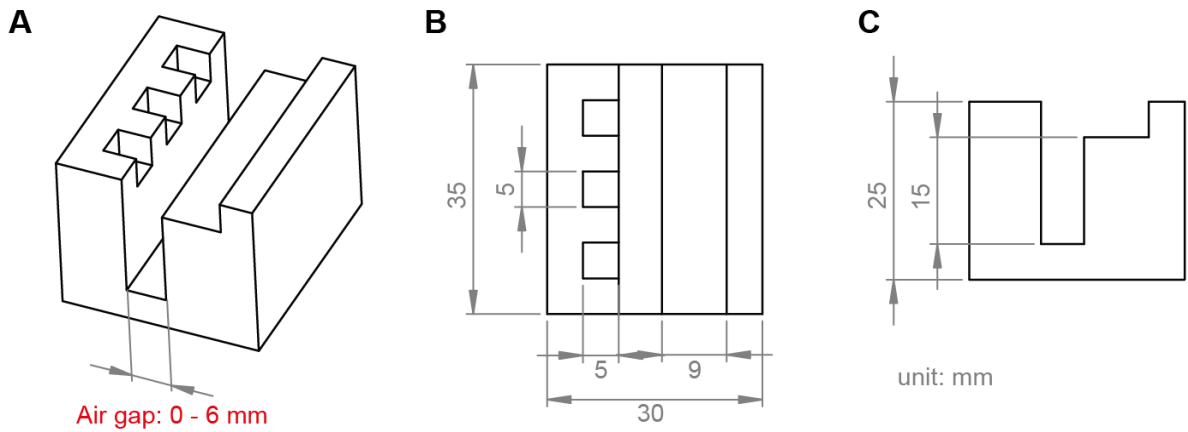

**Fig. S8. Sample holder used for bridging experiments.**
(**A**) 3D representation indicating the air gap that was changed to create a distance for the mycelium to bridge. (**B,C**) Drawings displaying different cross-sections of the sample holder.

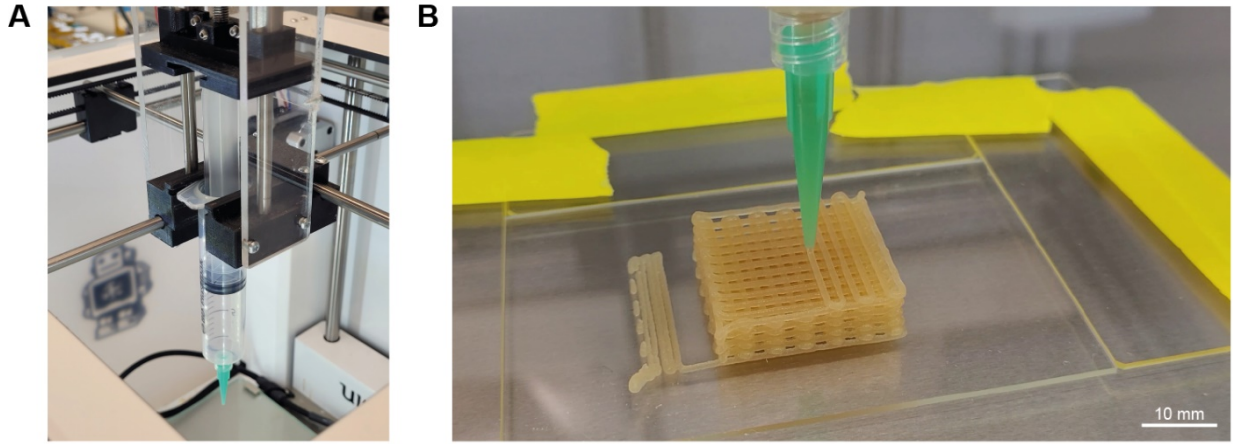

**Fig. S9. 3D Printing setup.**
(**A**) Custom-built syringe pump mounted on a fused filament fabrication printer (Ultimaker 2+) to create the direct ink writing setup (DIW). (**B**) DIW of the mycelium ink into a grid structure.

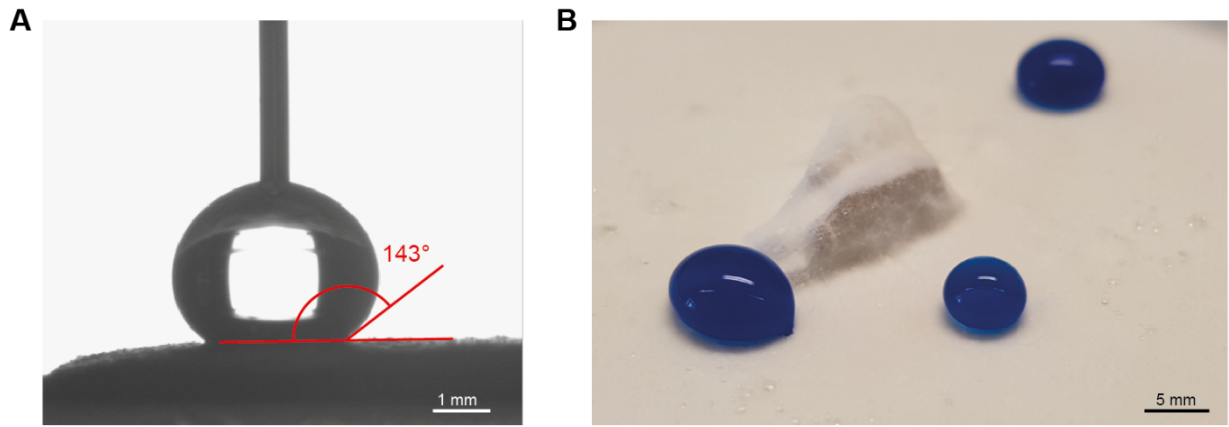

**Fig. S10. Surface hydrophobicity of printed structures.**
(**A**) Contact angle measurement on a mycelium-covered surface. (**B**) Colored water droplets on a plate covered with mycelium.

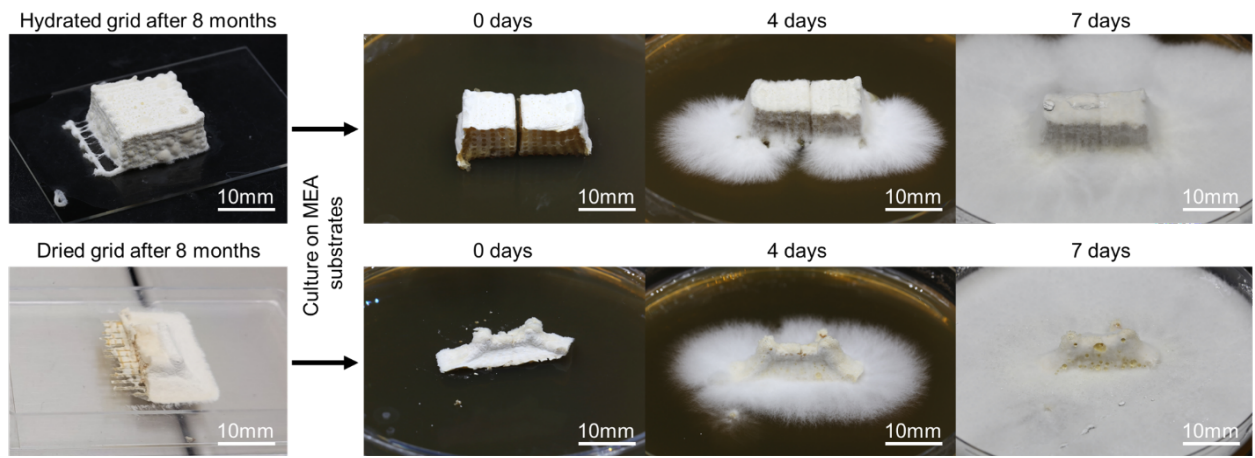

**Fig. S11. Long-term resilience and biological viability of mycelia-based structures.**

Printed living grids that were maintained hydrated (upper row) or dried (lower row) for 8 months are able to re-grow the mycelial network after culturing on malt extract growth plates. Both samples quickly regenerated new mycelium and colonized the plates within 7 days of growth.

**Movie S1. Mechanical robustness of mycelium-based living objects.**

The video shows the deformation of mycelium-covered objects under tensile forces, compressive load and under the action of a cutting wire. The wire cutting experiment demonstrates the toughness of the mycelium skin grown around the printed grids. The compression tests are also performed on samples with cut sides to reveal the deformation of the interior of the structure during mechanical loading.

**Movie S2. Growth of mycelium between printed filaments.**

The video shows a sequence of confocal microscopy images during the growth of mycelium between two printed filaments over a time window of 5 days. A 3D representation of the mycelium network is shown at the top, whereas the bottom displays a projection of the confocal images along the out-of-plane z-axis.

**Movie S3. Tensile testing of self-healing mycelium-covered structures.**

The video displays the deformation of sample A (Fig. 4E) when subjected to tensile loading. The first part shows the intact sample deformed until the first fracture event. The second part depicts the fracture of the same sample after self-healing.

**Movie S4. Fabrication and testing of mycelium-based skin for rolling robot.**

The footage first shows the deposition of the mycelium ink onto half-spherical mandrels using a 5-axis printer, followed by the growth of the mycelium to generate the living robotic skin. The video then displays the robustness of the skin and its ability to protect the robot from harsh environments, such as gravel, rocks and water.

**Movie S5. Fabrication and testing of mycelium-based skin for robotic gripper.**

The video illustrates the 3D printing of non-planar grid structures onto a mandrel with the shape of a robotic finger. After growth, assembly and fusion steps, the mycelium-covered gripper is used to deposit the rolling robot into a water bath. The robotic gripper retrieves the rolling robot again from the water after a short underwater motion. The hydrophobic skin covering the rolling robot and the gripper protects them against damage in water.